\documentclass[aps,prl,reprint,superscriptaddress]{revtex4-1}
\pdfoutput=1
\usepackage{easy-todo}
\usepackage{color}
\usepackage{graphicx}
\usepackage{float}
\usepackage{amsmath}
\usepackage{amsfonts}
\usepackage{amssymb}
\usepackage{microtype}
\usepackage{verbatim}
\usepackage{bbold}
\usepackage{hyperref}
\usepackage{cleveref}
\usepackage[normalem]{ulem} 
\allowdisplaybreaks
\usepackage{multirow}
\usepackage{bbm}
\usepackage{braket}
\hypersetup{
  colorlinks   = true, 
  urlcolor     = blue, 
  linkcolor    = blue, 
  citecolor   = blue 
}

\usepackage{textcomp}

\usepackage[smallerops]{newtxmath}

\DeclareMathOperator{\tr}{tr}

\begin{document}

\author{Pieter W. Claeys}
\email{claeys@pks.mpg.de}
\affiliation{Max Planck Institute for the Physics of Complex Systems, 01187 Dresden, Germany}
\affiliation{TCM Group, Cavendish Laboratory, University of Cambridge, Cambridge CB3 0HE, UK}

\author{Marius Henry}
\affiliation{ETH Z\"urich, 8092 Z\"urich, Switzerland}

\author{Jamie Vicary}
\affiliation{Computer Laboratory, University of Cambridge, Cambridge, CB3 0FD, UK}

\author{Austen Lamacraft}
\affiliation{TCM Group, Cavendish Laboratory, University of Cambridge, Cambridge CB3 0HE, UK}

\title{Exact dynamics in dual-unitary quantum circuits with projective measurements}

\begin{abstract}
Dual-unitary circuits have emerged as a minimal model for chaotic quantum many-body dynamics in which the dynamics of correlations and entanglement remains tractable. Simultaneously, there has been intense interest in the effect of measurements on the dynamics of quantum information in many-body systems. In this work we introduce a class of models combining dual-unitary circuits with particular projective measurements that allow the exact computation of dynamical correlations of local observables, entanglement growth, and steady-state entanglement. We identify a symmetry preventing a measurement-induced phase transition and present exact results for the intermediate critical purification phase.
\end{abstract}

\maketitle
\emph{Introduction.} -- Understanding the dynamics of quantum many-body systems remains a crucial problem with applications ranging from condensed matter physics to quantum information. Computing dynamical quantities such as correlation functions and entanglement growth is a notoriously hard problem in interacting systems, both numerically and analytically. In recent years unitary circuits have moved beyond models of quantum computation to minimal models for the study of general unitary dynamics governed by local interactions \cite{nahum_quantum_2017,khemani_operator_2018,von_keyserlingk_operator_2018,nahum_operator_2018,chan_solution_2018,rakovszky_sub-ballistic_2019,friedman_spectral_2019,garratt_local_2021}. A special class of such circuits, termed \emph{dual-unitary circuits}, remains amenable to exact calculations \cite{bertini_exact_2019,gopalakrishnan_unitary_2019}. These circuits are characterized by an underlying spacetime-duality, resulting in unitary dynamics in both time and space. This duality allows for exact calculations of the dynamics of correlation functions of local observables \cite{bertini_exact_2019,piroli_exact_2020,gutkin_exact_2020,claeys_ergodic_2021,kos_correlations_2021}, out-of-time-order correlators \cite{claeys_maximum_2020,bertini_scrambling_2020}, entanglement \cite{gopalakrishnan_unitary_2019,bertini_entanglement_2019}, indicators of quantum chaos \cite{bertini_exact_2018,bertini_random_2021,kos_chaos_2021,fritzsch_eigenstate_2021}, and dual-unitary circuits have since been the subject of active theoretical \cite{rather_creating_2020,aravinda_dual-unitary_2021,prosen_many-body_2021,jonay_triunitary_2021,borsi_remarks_2022,gombor_superintegrable_2022,suzuki_computational_2022,lerose_influence_2021,giudice_temporal_2021,reid_entanglement_2021,lerose_overcoming_2022, zabalo_operator_2022,ho_exact_2021,claeys_emergent_2022,ippoliti_dynamical_2022,rather_construction_2022,kasim_dual_2022} and experimental studies \cite{chertkov_holographic_2021,mi_information_2021}.

Moving beyond the purely unitary dynamics of closed quantum systems, circuit models also present a natural playground for nonunitary dynamics by introducing projective measurements at given points in spacetime. As the measurement rate is increased, such systems may undergo a transition from volume-law to area-law scaling of steady-state entanglement with subsystem size: the celebrated measurement-induced phase transition (MIPT)
 \cite{li_quantum_2018,skinner_measurement-induced_2019,chan_unitary-projective_2019,li_measurement-driven_2019,gullans_dynamical_2020,potter_entanglement_2021,zabalo_infinite-randomness_2022}.

\begin{figure}[htb!]
\includegraphics[width=\columnwidth]{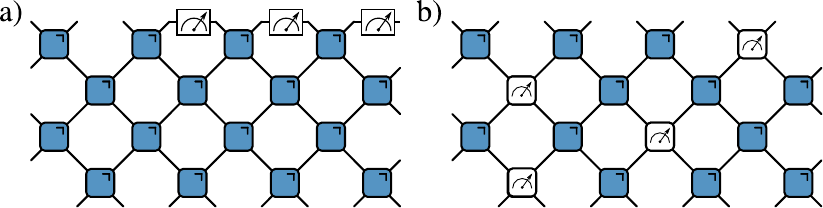}
\caption{a) Dual-unitary circuit followed by measurements. b) Dual-unitary circuit interspersed with projective measurements. 
\label{fig:intro}
\vspace{-1\baselineskip}}
\end{figure}

In general, including measurements in dual-unitary circuits destroys their exact solvability. Several recent works have shown, however, that when dual-unitary dynamics is \emph{followed} by projective measurements in specific measurement bases (see Fig.~\ref{fig:intro}a)), the resulting circuit remains amenable to theoretical analysis. This approach was used to study higher notions of thermalization following projective measurements on a bath  \cite{ho_exact_2021,claeys_emergent_2022,ippoliti_dynamical_2022}. 
In this work we introduce a class of circuits where dual-unitary dynamics is combined with projective measurements at \emph{arbitrary} locations, as in models displaying the MIPT (see Fig.~\ref{fig:intro}b)). With particular restrictions on the measurement basis and some of the gates, we are able to exactly calculate the dynamics of correlations and entanglement growth.

Recent works have used the idea of spacetime duality to study nonunitary dynamics using unitary circuits \cite{ippoliti_postselection-free_2021,ippoliti_fractal_2022,lu_spacetime_2021}. In these papers the nonunitary dynamics is interpreted as arising from measurements with a fixed set of outcomes, and the unitary dynamics in the spatial direction provides a way to avoid this postselection. In contrast, our focus is on the dynamics of circuits with arbitrary measurement outcomes.

In dual-unitary circuits all correlations vanish everywhere except at the edge of the causal light cone, where they can be efficiently calculated using a quantum channel approach \cite{bertini_exact_2019,claeys_maximum_2020}. We show that this approach extends to our hybrid circuits. The solvability is preserved by (i) performing measurements in a unitary basis and (ii) restricting the gates on the edge of the measurement light cones such that no new correlations are created. For a fixed measurement rate the calculation of the average entanglement growth can be mapped to a toy model previously studied by Ippoliti and Khemani \cite{ippoliti_postselection-free_2021}, where it described circuits of gates of a particularly simple form. We present exact expressions for the resulting steady-state entanglement as well as the entanglement growth -- which is shown to remain close to the maximal entanglement growth of dual-unitary circuits. In Ref.~\cite{ippoliti_postselection-free_2021} numerical studies and qualitative arguments were used to argue that the toy model does not support a measurement-induced phase transition, but is rather described by a critical purification phase for all measurement rates $0<p<1$. Here we identify a hidden symmetry in the expression for the entanglement that relates measurement rates $p$ and $1-p$ and prevents such a phase transition. We derive an analytic expression for the averaged entropy $\overline{S}$ [Eq.~\eqref{eq:exact_S}]. For intermediate measurement rates this entropy grows linearly in time at short times, $\overline{S} \propto t$, and at long times it scales with subsystem size $\ell$ as $\overline{S}\propto\ell^{1/2}$.

\begin{figure*}[t] 
\includegraphics[width=0.9\textwidth]{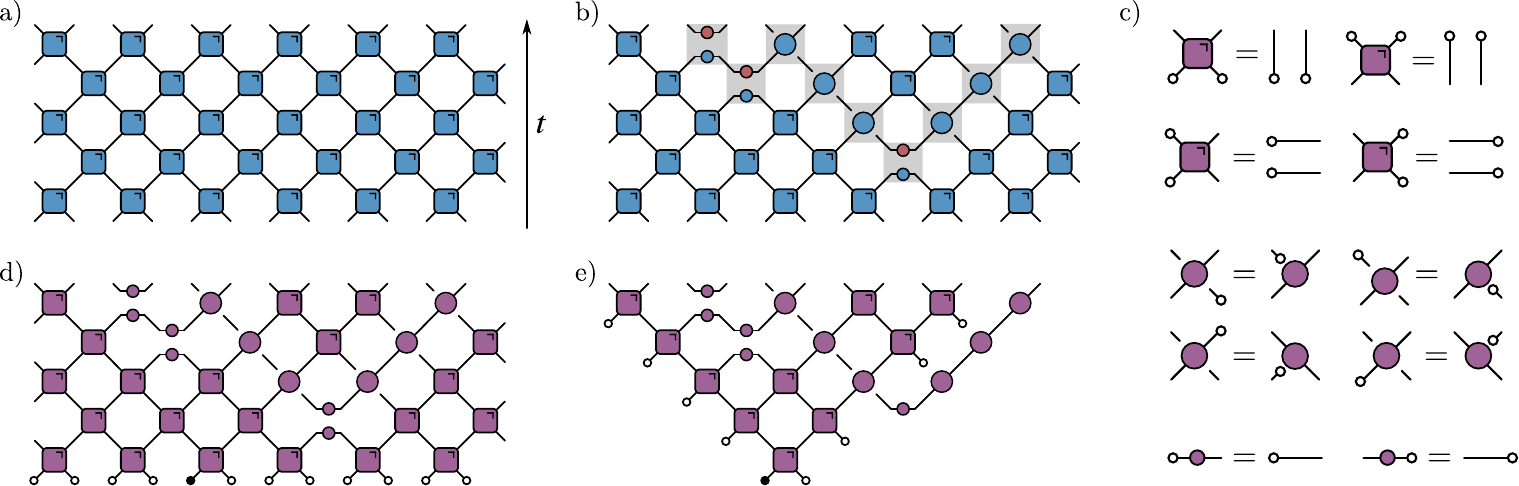}
\caption{a) Unitary circuit without measurements. The lattice sites are arranged in a horizontal row with two-site gates acting alternately on the even and odd links between them. Discrete time runs vertically. b) Hybrid nonunitary ciruit where projective measurements in a UEB have been introduced, with restricted dual-unitary gates on the edges of the measurement light cones. Both the measurements and their light cones have been highlighted. Note that measurements can occur on the light cone of previous measurements. c) Graphical identities in the folded representation. d) Time-evolved local operator. e) Time-evolved local operator simplified using the identities from c).
\label{fig:circuits}}
\vspace{-1\baselineskip}
\end{figure*}    

\emph{Hybrid circuits.} -- We consider circuits where the unitary dynamics is generated by two-site unitary gates, with each gate $U$ and its hermitian conjugate graphically represented as
\begin{align}\label{eq:def_U_Udag}
U_{ab,cd} = \vcenter{\hbox{\includegraphics[width=0.15\linewidth]{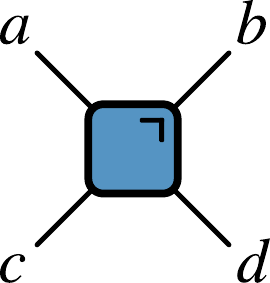}}},\qquad U^{\dagger}_{ab,cd} =  \vcenter{\hbox{\includegraphics[width=0.15\linewidth]{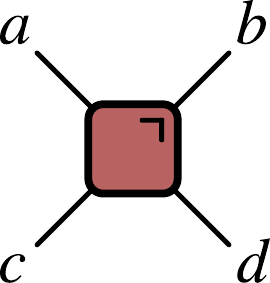}}}. 
\end{align}
In this graphical notation each leg carries a local $q$-dimensional Hilbert space, and the indices of legs connecting two operators are implicitly summed over (see e.g. Ref.~\cite{orus_practical_2014}). The dual of $U$ is defined as $\tilde{U}_{ab,cd} = U_{db,ca}$, and dual-unitarity is defined by demanding that both $U$ and $\tilde{U}$ are unitary \cite{bertini_exact_2018},
\begin{align}
U U^{\dagger} = U^{\dagger} U = \mathbbm{1}, \qquad \tilde{U} \tilde{U}^{\dagger} = \tilde{U}^{\dagger} \tilde{U} = \mathbbm{1}\,.
\end{align}
The projective dynamics consists of two-site measurements in a unitary error basis (UEB), e.g. the Bell pair basis \footnote{For $q=2$ all unitary error bases are equivalent to the Pauli matrices, for which the corresponding state are given by the Bell pair states.} \cite{knill_non-binary_1996}. Such a basis consists of states $\ket{\alpha_n}=\sum_{i,j=1}^q\left(\alpha_n\right)_{ij}\ket{i,j}$, parametrized by a collection of unitary matrices $\{\alpha_n,n=1\dots q^2\}$ satisfying the orthogonality property $\tr(\alpha_n^{\dagger}\alpha_m) = q\delta_{m,n}$. Because of this orthonormality the corresponding states form a complete basis for the $q^2$-dimensional two-site Hilbert space, which also underlies their use in quantum teleportation \cite{knill_non-binary_1996,reutter_biunitary_2019}. Graphically, 
\begin{align}
|\alpha_n \rangle_{ab}  \langle \alpha_n|_{cd} = \vcenter{\hbox{\includegraphics[width=0.15\linewidth]{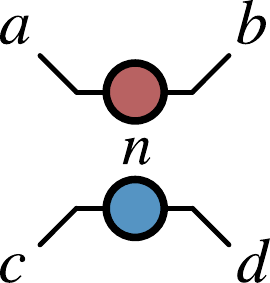}}}\,,
\end{align}
acting as a projector in the time direction and as the product of one-site unitaries in the spatial direction \footnote{Note that these states are normalized as $\braket{\alpha_n|\alpha_m} = q \delta_{m,n}$, which will be important when considering normalization following projection.}.

An evolution operator $\mathcal{U}(t)$ for an infinite lattice can be obtained by considering the `brick-wall' quantum circuits illustrated in Fig.~\ref{fig:circuits}, where the number of discrete time steps corresponds to the number of rows of unitary gates. For dual-unitary gates, these circuits are unitary along both the time and space direction.
Measurements can naively be introduced by replacing two-site dual-unitary gates by measurements in a UEB. For any measurement outcome the corresponding nonunitary circuit remains unitary along the spatial direction, returning a specific circuit that is spacetime dual to a unitary circuit (of the form studied in Refs.~\cite{ippoliti_postselection-free_2021,ippoliti_fractal_2022,lu_spacetime_2021}). However, the correlations and entanglement dynamics in such a system are no longer tractable. In order to obtain circuits amenable to exact calculations, it is necessary to restrict the unitary gates on the edge of the measurement light cones, as illustrated in Fig.~\ref{fig:circuits}b). We restrict these gates to be of the form $U = S \left[u_1 \otimes u_2\right]$, with $S$ a swap gate, $S_{ab,cd}=\delta_{ad}\delta_{bc}$, and $u_{1,2} \in U(q)$ one-site unitary gates. We introduce a short-hand graphical notation,
\begin{align}
\vcenter{\hbox{\includegraphics[width=0.15\linewidth]{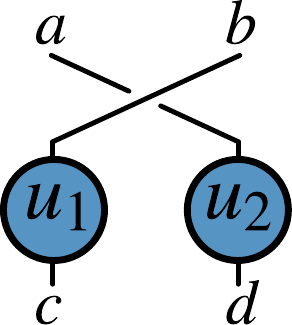}}}\equiv \vcenter{\hbox{\includegraphics[width=0.15\linewidth]{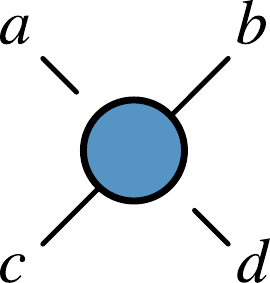}}} = (u_1)_{bc}(u_2)_{ad}\,.
\end{align}
These gates correspond to a restricted set of dual-unitary gates \cite{claeys_ergodic_2021}, but the remaining dual-unitary gates in the circuit can be chosen freely. This circuit is directly inspired by biunitary circuits \cite{reutter_biunitary_2019,biunitary_2022}.

\emph{Correlation functions.} -- We illustrate the dynamics of correlations for infinite-temperature correlation functions of the form
\begin{equation}\label{eq:def_corr}
c_{\rho\sigma}(x,t) = \tr\left[ \mathcal{U}^{\dagger}(t)\rho (0) \mathcal{U}(t) \sigma(x) \right]  / \tr(\mathbbm{1}),
\end{equation}
where $\rho, \sigma \in \mathbbm{C}^{q\times q}$ are operators acting on the local $q$-dimensional Hilbert space, and $\rho(x), \sigma(x)$ act as $\rho,\sigma$ on site $x$ and as the identity everywhere else. We choose $\tr(\rho)=1$ and $\sigma$ traceless without loss of generality. In order to describe the correlation dynamics, we introduce the commonly used `folded' notation, mapping $U^{\dagger} \otimes U^T$ to gates acting in a doubled Hilbert space,
\begin{align}
\vcenter{\hbox{\includegraphics[width=0.74\linewidth]{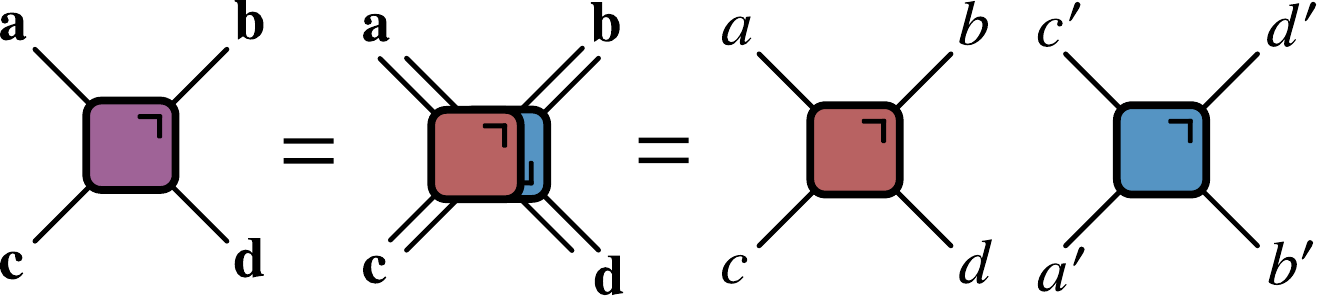}}}\,
\end{align}
with $\mathbf{a}=(a,a')$ etc., and similar for projectors and dressed swap gates, and mapping operators to states in this doubled Hilbert space. We represent the (normalized) folded identity by a white circle and general operators $\sigma$ by filled circles, 
\begin{align}
\vcenter{\hbox{\includegraphics[width=0.03\linewidth]{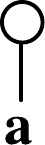}}} = \frac{1}{\sqrt{q}}\mathbbm{1}_{a,a'} = \frac{\delta_{a,a'}}{\sqrt{q}},\quad \vcenter{\hbox{\includegraphics[width=0.055\linewidth]{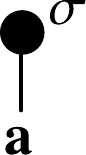}}} = \mathbbm{\sigma}_{a,a'} \, \,.
\end{align}
From the combined unitarity and dual-unitarity, the building blocks of these hybrid circuits satisfy a set of graphical identities illustrated in Fig.~\ref{fig:circuits}c). 

First we note that correlation functions on the lightcone, i.e. $x=\pm t$, can be evaluated in the exact same way as for general unitary circuits \cite{claeys_maximum_2020}. These functions are graphically represented as
\begin{align}
c_{\rho\sigma}(x=t,t) = \vcenter{\hbox{\includegraphics[width=0.7\linewidth]{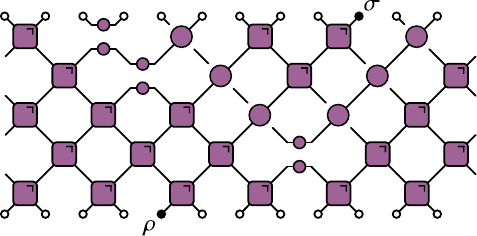}}}\,,
\end{align}
where we have chosen a representative example. The two nontrivial operators (black circles) are separated by a distance equal to the number of time steps. Using the identities from Fig.~\ref{fig:circuits}, almost all gates and measurements can be removed to return
\begin{align}
c_{\rho\sigma}(x=t,t) = \vcenter{\hbox{\includegraphics[width=0.45\linewidth]{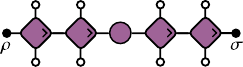}}}\,.
\end{align}
Such expressions can be exactly evaluated at arbitrary large times by interpreting the contracted folded gates as quantum channels \cite{bertini_exact_2019,claeys_ergodic_2021}. Generically, these result in exponentially decaying correlations and ergodicity, but fine-tuned dual-unitary circuits can lead to richer light-cone dynamics \cite{bertini_exact_2019,claeys_ergodic_2021,aravinda_dual-unitary_2021}.

Before moving on to general correlation functions, note that for $\sigma=\mathbbm{1}$ this expression returns the normalization of the state after the projective measurements. This diagram can be directly evaluated to return $\tr(\rho)=1$. Remarkably, for our circuit construction the above expression is already normalized, indicating that all measurement outcomes are equally likely. Note that this equality always holds in circuits constructed out of dual-unitary gates and measurements in a UEB -- however, other measurements probabilities generally depend on the measurement outcome. 

For general correlation functions we first note that the time-evolved operator $ \mathcal{U}^{\dagger}(t)\rho (0) \mathcal{U}(t)$ can be graphically represented as in Figs.~\ref{fig:circuits}d) and e), where it is straightforward to check that all dual-unitary gates outside of the light cone $x=\pm t$ can be removed using the above identities. 
Outside of the causal light cone this operator only acts nontrivially if the site $x$ is located on the light cone of some measurement. However, for the illustrated diagram the infinite-temperature correlation functions can be directly shown to vanish at these sites. Any contraction on the edge of the measurement light cone can be propagated through the dressed swap gates, such that the resulting correlation function will always contain a factor $\tr(\sigma)$ and hence vanish since
\begin{align}
\vcenter{\hbox{\includegraphics[width=0.65\linewidth]{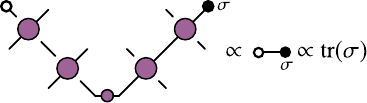}}} = 0\,.
\end{align}
Correlation functions outside the causal light cone here vanish. Using the same arguments as for purely dual-unitary circuits (see e.g. Ref.~\cite{bertini_exact_2019}), the unitarity along the spatial direction implies that correlations also vanish inside the light cone. The only nontrivial correlations are those on the edge of the light cone, $x=\pm t$, and these can be calculated in the same way as for regular dual-unitary circuits. 
It is possible to generate nontrivial correlations outside the light cone through repeated measurements on the original light cone, in which case correlation functions will vanish everywhere except on the edge of a shifted light cone, where they can again be calculated using the same quantum channels and hence with the same computational cost (see Supplemental Material \footnote{Supplemental Material.}).

\emph{Entanglement dynamics.} -- Next to their strict light cone, dual-unitary circuits are also characterized by their maximal entanglement growth \cite{zhou_maximal_2022}. Whereas initial results on entanglement dynamics were restricted to the kicked Ising model at the self-dual point \cite{gopalakrishnan_unitary_2019}, it was later shown that the time evolution of the entanglement entropy of a connected block can be explicitly characterized for a special class of initial `solvable states' preserving the dual-unitarity \cite{piroli_exact_2020}. The simplest case of such a solvable state is given by
\begin{equation}
\ket{\Psi_0} \propto \bigotimes_{n} \left(\sum_{i=1}^q\ket{i,i}_{2n,2n+1}\right),
\end{equation}
reducing to the state where all neigboring states are prepared in a Bell pair $(\ket{00}+\ket{11})/\sqrt{2}$ for local qubits ($q=2$). Such initial states are also required when considering dynamics that is the spacetime dual of a unitary circuit \cite{ippoliti_postselection-free_2021,ippoliti_fractal_2022}. We choose a subsystem $A$ to consist of $\ell$ consecutive sites in the infinite lattice, leading to a reduced density matrix
\begin{align}
\rho_A(t) = \tr_{\overline{A}}\left[\mathcal{U}(t)\ket{\Psi_0}\bra{\Psi_0}\mathcal{U}(t)^\dagger\right],
\end{align}
tracing out the complement of $A$. In order to measure the entanglement entropy, we study the growth of R\'enyi entropies,
\begin{equation}
S_{A}^{(\alpha)}(t)= \frac{1}{1-\alpha} \log \tr\left[(\rho_A(t))^{\alpha}\right],\quad \alpha > 0\,.
\end{equation}
In dual-unitary circuits the infinite bath $\overline{A}$ acts as a perfectly Markovian bath, i.e. a `perfect dephaser' \cite{lerose_influence_2021}, and this property also holds for the proposed class of hybrid circuits (see Supplemental Material \cite{Note3}). This property implies that the reduced density matrix after $t$ time steps can be represented by
\begin{equation}
\rho_A(t) = \vcenter{\hbox{\includegraphics[width=0.6\linewidth]{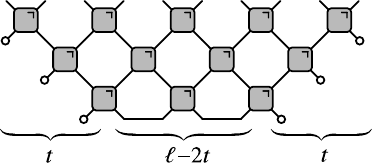}}}
\end{equation}
where every gate now represents either a folded gate or measurement,
\begin{align}
\vcenter{\hbox{\includegraphics[width=0.10\linewidth]{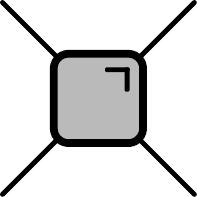}}} = \vcenter{\hbox{\includegraphics[width=0.10\linewidth]{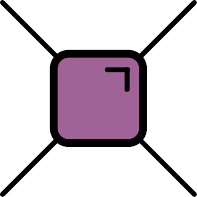}}} \quad\textrm{or}\quad\vcenter{\hbox{\includegraphics[width=0.10\linewidth]{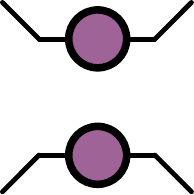}}} \quad\textrm{or}\quad\vcenter{\hbox{\includegraphics[width=0.10\linewidth]{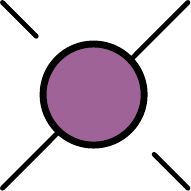}}}\, ,
\end{align}
depending on the layout of the original circuit. For our given circuit constructions every realization of dual-unitary gates and measurements will lead to a reduced density matrix that is a unitary transformation of a projector times the (properly normalized) identity. These models have a flat entanglement spectrum, with all eigenvalues of $\rho_A(t)$ either zero or degenerate \cite{Note3}. This property guarantees that the R\'enyi entropies are independent of the R\'enyi index $\alpha$ and equal the von Neumann entropy $S_A$. After times $t > \ell/2$, the reduced density matrix can be written as
\begin{equation}
\rho_A(t \geq \ell/2) = \vcenter{\hbox{\includegraphics[width=0.45\linewidth]{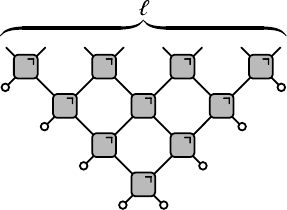}}}
\end{equation}
This representation can be used to calculate the average entropy for a hybrid circuit where every gate is either a projective measurement with probability $p$ or a dual-unitary gate with probability $1-p$ (again with restricted dual-unitary gates on the light cone of every measurement), chosen independently. If a MIPT occurs, this is exactly the quantity in which it is expected to be apparent. As argued in the Supplemental Material \cite{Note3}, the entanglement entropy $\overline{S_{A}}$ averaged over different measurement realizations can be expressed as 
\begin{align}\label{eq:diagram_Sa}
\vcenter{\hbox{\includegraphics[width=0.45\linewidth]{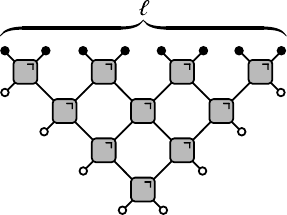}}} = 1 + \epsilon\, \overline{S_A(t)} + \mathcal{O}(\epsilon^2)\,,
\end{align}
with
\begin{align}
\vcenter{\hbox{\includegraphics[width=0.55\linewidth]{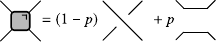}}}\,,
\end{align}
and boundaries satisfying $\vcenter{\hbox{\includegraphics[width=0.06\linewidth]{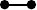}}}=\vcenter{\hbox{\includegraphics[width=0.06\linewidth]{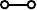}}}=1$ and $\vcenter{\hbox{\includegraphics[width=0.06\linewidth]{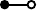}}}=1-\epsilon \ln(q)$. These boundary conditions effectively count the total number of nonzero eigenvalues. The expansion in $\epsilon$ is required such that the product of overlaps returns a summation where every nontrivial overlap contributes one bit of entropy. Since the entanglement is purely set by the number of nonzero eigenvalues, and this number is independent of the specific choice of dual-unitary gates and measurement outcomes, these gates can be replaced by the simplest choice of swap gates and projections on the identity respectively (even if the dynamics of correlations functions is typically different from those of such a simplified circuit). The entanglement dynamics for such a circuit was previously studied numerically in Ref.~\cite{ippoliti_postselection-free_2021} as a toy model, and here naturally arises as an effective model for the entanglement growth in our hybrid circuits. 
We emphasize that the unitary and projective dynamics in these hybrid dual-unitary models can be much more complicated than those of the toy model, but the entanglement entropies will be identical (same as in purely dual-unitary dynamics, see e.g. Ref.~\cite{gopalakrishnan_unitary_2019}). 

\begin{figure}[tb!]
\includegraphics[width=\columnwidth]{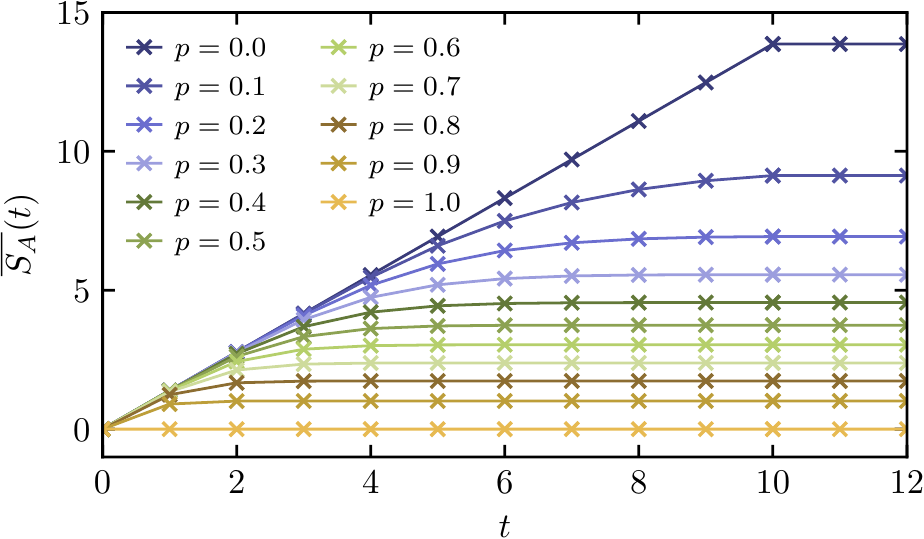}
\caption{Growth of the averaged entanglement entropy as function of time for different measurement rates $p$ and subsystem size $\ell=20$. 
\label{fig:dyn_S}}
\end{figure}

\begin{figure}[tb!]
\includegraphics[width=\columnwidth]{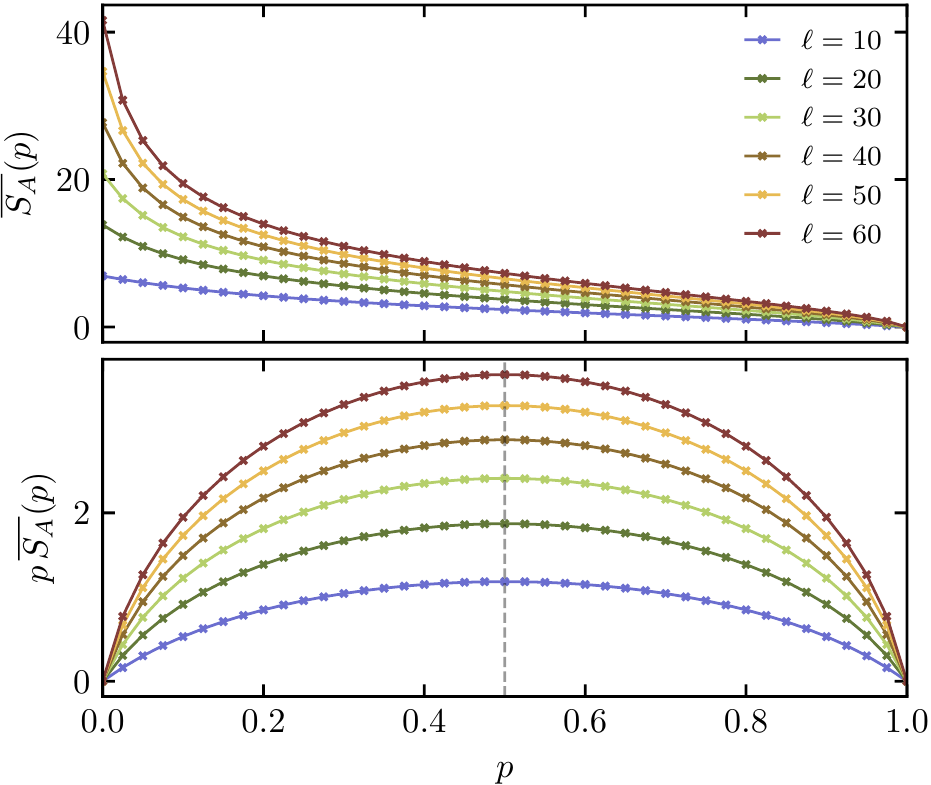}
\caption{Top: Averaged entanglement entropy $\overline{S_A}$ for different measurement rates $p$ and subsystem sizes $\ell$. Bottom: Entanglement entropy rescaled by the measurement probability. Vertical dashed line denotes the symmetric point $p=1/2$.
\label{fig:static_S}}
\end{figure}

Numerical results are presented in Fig.~\ref{fig:dyn_S} for the averaged entanglement growth for different measurement rates $p$ in a subsystem with a fixed number of sites $\ell=20$. At short times the entanglement growth remains linear, with a slope given by the entanglement velocity, which remains close to the maximal velocity of dual-unitary circuits even for a nonvanishing measurement rate. After a single time step
\begin{align}
\overline{S_A}(t=1) = (1-p^{\ell/2})\, 2\ln(q)\,,
\end{align}
and, more generally,
\begin{align}
\overline{S_A}(t) \approx 2 \ln(q)\, t \quad \textrm{if} \quad t \ll \ell/2,
\end{align}
up to corrections that are exponentially small in $\ell$ \cite{Note3}, before reaching a steady-state value $\overline{S_A}(t\to\infty)$.
This averaged steady-state entanglement is shown in Fig.~\ref{fig:static_S} for different system sizes $\ell$ and measurement rates $p$. Two simple limits can be clearly identified: for $p=0$ we recover the volume-law entanglement $\overline{S_A} = \ln(q)\, \ell$ of dual-unitary circuits, and for $p=1$ no entanglement growth is possible and $\overline{S_A}(t)=0$. 

Ref.~\cite{ippoliti_postselection-free_2021} argued that no phase transition occurs between volume-law and area-law scaling for intermediate measurement rates. Rather, the intermediate phase corresponds to a critical purification phase with intermediate entanglement scaling $\propto\sqrt{\ell}$.
In Ref.~\cite{ippoliti_postselection-free_2021}, this scaling was deduced from the diffusive dynamics of qubits in this toy model. We now show how these results can be obtained in our model by explicit calculation.

\begin{figure}[tb!]
\includegraphics[width=\columnwidth]{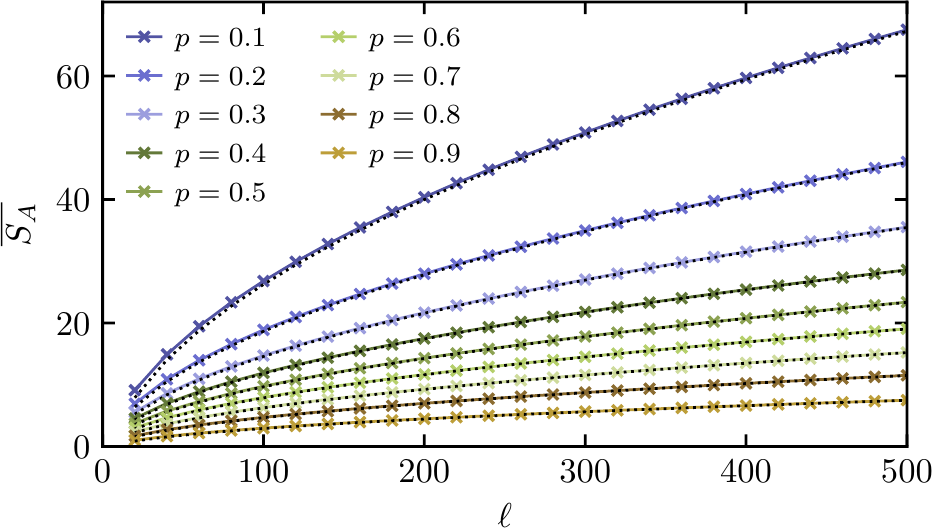}
\caption{Averaged steady-state entanglement entropy as function of subsystem size for different measurement rates $p$. Full lines denote exact results, dashed lines denote the asymptotic expansion. 
\label{fig:scaling_S}}
\end{figure}

First, the absence of a measurement-induced phase transition can be demonstrated from a previously unobserved symmetry: the averaged steady-state entanglement $\overline{S_A}(p)$ for fixed measurement rate $p$ satisfies
\begin{align}
p \overline{S_A}(p) = (1-p)\overline{S_A}(1-p).
\end{align}
In other words, the entanglement scaling with subsystem size is identical for measurement rates $p$ and $1-p$, only with a different prefactor, preventing volume-law scaling for small $p$ and area-law scaling for large $p$. This symmetry is illustrated in Fig.~\ref{fig:static_S} and can be established from the graphical representation \eqref{eq:diagram_Sa}. This diagram can be explicitly evaluated in terms of paths connecting different sites to return \cite{Note3}
\begin{align}
&p \overline{S_A}(p)/\ln(q) = p(1-p)\ell \nonumber\\
&\quad-2\sum_{n=1}^{\ell/2-1} \left(\frac{\ell}{2}-n\right)\sum_{k=1}^n N(n,k) p^{2k}(1-p)^{2(n-k+1)}\,,
\end{align}
with $N(n,k)=\frac{1}{n} {n \choose k} {n\choose k-1}$ the Narayana numbers. These satisfy $N(n,k) = N(n,n-k+1)$, such that the above expression is explicitly symmetric under exchange of $p$ and $1-p$. A closed-form expression can be obtained in terms of the Gaussian hypergeometric function $\,_2F_1$ \cite{Note3},
\begin{align}\label{eq:exact_S}
\overline{S_A}(p)&=\frac{\ln(q)}{p}\left[\,_2F_1(-\ell/2,-1/2,1;4p(1-p))-1\right] \\
&\approx \frac{\ln(q)}{p}\left[\sqrt{\frac{8p(1-p)\,\ell}{\pi}}-1\right]\,,
\end{align}
where the approximation denotes the asymptotics for large $\ell$ \cite{bateman_manuscript_project_higher_1953}. This result analytically establishes an exact $\sqrt{\ell}$ scaling of the entanglement entropy with subsystem size for any intermediate measurement rate $p\neq 0,1$. At the symmetric point $p=1/2$ the above expression simplifies to
\begin{align}
\frac{\overline{S_{A}}(p=1/2)}{2\ln(q)} = \frac{\ell+1}{2^\ell}{\ell \choose \ell/2}-1 \approx \sqrt{\frac{2\ell}{\pi}}-1\,,
\end{align}
with the large-$\ell$ approximation reproducing Stirling's formula. The exact result \eqref{eq:exact_S} is compared with the asymptotic formula in Fig.~\ref{fig:scaling_S}, showing good agreement. The apparently slow entanglement growth for large values of $p$ is clearly due to a smaller prefactor, but exhibits the same scaling with $\ell$ everywhere.

\emph{Conclusion.} -- In this work we have introduced a class of nonunitary circuits that combine dual-unitary dynamics with projective measurements in such a way that the exact solvability of dual-unitary circuits is preserved. Projective measurements are performed in a unitary error basis, preserving unitarity along the spatial direction, and the dual-unitary gates on the edge of the light cone following each measurement are restricted to dressed swap-gates. This construction guarantees 1) that the only nonvanishing dynamical correlation functions are the light-cone correlation functions, which can be calculated using the same quantum channel approach as for dual-unitary circuits, and 2) that for initial solvable states the reduced density matrices have a flat entanglement spectrum. This latter property allows for the exact calculation of averaged entanglement entropies for fixed measurement rates. We recover the limits of volume-law and area-law scaling and find that no measurement-induced phase transition is observed as the measurement rate is varied, but the system rather enters a critical purification phase. By identifying an underlying symmetry we analytically established the absence of any transition and calculated the steady-state entanglement, scaling as $\sqrt{\ell}$ with subsystem size $\ell$. These hybrid circuits require neither randomness in the unitary gates, a large local Hilbert space, or restrictions to Clifford gates, and are solvable for any choice of dual-unitary gates, unitary error basis, and dimension of the local Hilbert space.
Dual-unitary circuits have served as a powerful analytical tool in the study of unitary quantum dynamics in recent years, and these hybrid nonunitary circuits can serve a similar role for nonunitary dynamics. 

\emph{Acknowledgements.} -- P.W.C. and A.L. gratefully acknowledge support from EPSRC Grant No. EP/P034616/1. J.V. gratefully acknowledges funding from the Royal Society.

\bibliography{Library.bib}

\newpage
\begin{widetext}
\section*{Supplemental Material: Exact dynamics in dual-unitary quantum circuits with projective measurements}

\section{Light-cone correlation functions}
In this Section we detail the quantum channel construction for correlations functions for completeness. This construction is fully analogous to the quantum channel approach for fully unitary dual-unitary circuits (see e.g.~\cite{bertini_exact_2019,claeys_maximum_2020}), with the only difference being that the quantum channels associated with either measurements or the light cones of measurements need to be restricted. Consider the circuit from the main text, for which the considered light-cone correlations can be graphically evaluated as
\begin{align}
\vcenter{\hbox{\includegraphics[width=0.55\linewidth]{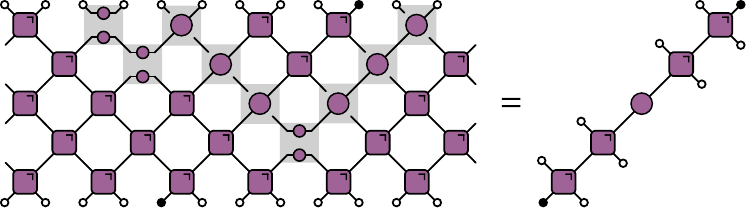}}} \,.
\end{align}
This representation holds irrespective of the specific choice of gates. For convenience, we now choose all dual-unitary gates to be identical. We similarly choose all dressed swap gates identical and parameterized as $S[u \otimes v]$. The action of the quantum channel $\mathcal{M}$ associated with the dual-unitary gate $U$ can be written as
\begin{align}
    \mathcal{M}(\rho) = \vcenter{\hbox{\includegraphics[width=0.07\linewidth]{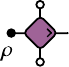}}} =\frac{1}{q}\,\vcenter{\hbox{\includegraphics[width=0.06\linewidth]{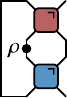}}} = \frac{1}{q}\tr\left[U^{\dagger}(\rho \otimes \mathbbm{1})U\right],
\end{align}
where in the second equality we have unfolded all gates. These are completely positive and trace-preserving maps satisfying $\mathcal{M}(\mathbbm{1})=\mathbbm{1}$, acting as a unital quantum channel \cite{bertini_exact_2019}. For a local $q$-dimensional Hilbert space these channels are operators acting on a $q^2$-dimensional space, such that their eigenvalue decomposition can be efficiently obtained and these correlation functions can be evaluated at arbitrary long times \cite{bertini_exact_2019}. 
The quantum channels set by dressed swap gates simplify to unitary transformations of $\rho$, e.g. for propagation along the right edge of the light cone,
\begin{align}
    \mathcal{S}(\rho) = \vcenter{\hbox{\includegraphics[width=0.07\linewidth]{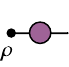}}} =\,\vcenter{\hbox{\includegraphics[width=0.06\linewidth]{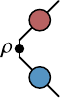}}}  = v^{\dagger}\rho v \,. 
\end{align}
The above correlation function then evaluates to $\tr[\sigma \mathcal{M}^2 \mathcal{S}\mathcal{M}^2(\rho)]$.

Nonzero correlations outside the light cone can similarly be evaluated using the same quantum channels. The only possibility for a nonzero correlation outside the light-cone is to have a measurement on the intersect of the light cone of the initial operator and the light cone of another measurement, leading to circuits of the form
\begin{align}
\vcenter{\hbox{\includegraphics[width=0.6\linewidth]{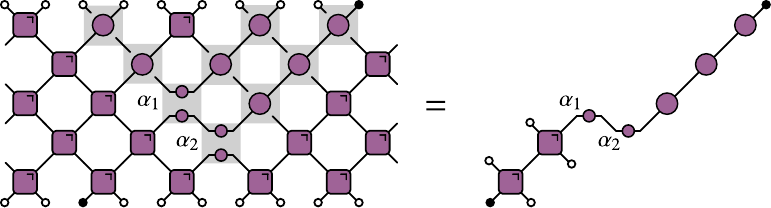}}} \,.
\end{align}
The quantum channel for a measurement can again be represented as a quantum channel corresponding to a unitary transformation, now with the unitary transformation depending on the measurement outcome. The ket from a projective measurement outcome $\ket{\alpha_n}$ results in a unitary transformation with $\alpha_n$, whereas the bra from a projective measurement outcome results in a unitary transformation with $\alpha_n^{\dagger}$,
\begin{align}
    \mathcal{S}_n(\rho)=\alpha_n^{\dagger} \rho \alpha_n, \qquad \mathcal{S}_n^{\dagger}(\rho) = \alpha_n \rho \alpha_n^{\dagger}\,.
\end{align}
The above correlation function now evaluates to $\tr[\sigma \mathcal{S}^3\mathcal{S}_2^{\dagger}\mathcal{S}_1\mathcal{M}^2(\rho)]$, depending on both measurement outcomes. Correlations at different sites vanish identically, such that the effect of the measurements is simply to shift the causal light cone.

\section{Derivation of the entanglement entropy}
For any hybrid circuit the reduced density matrix for a subsystem consisting of $\ell$ sites in an infinite system after $t$ time steps can be represented as
\begin{align}
\vcenter{\hbox{\includegraphics[width=0.55\linewidth]{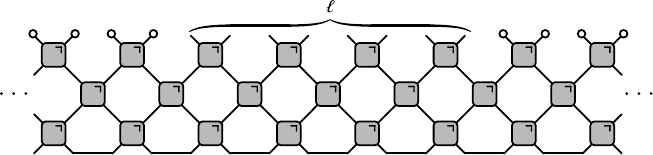}}}
\end{align}
here illustrated for $t=3$ and $\ell=8$. All gray gates are either dual-unitary gates or projections in a UEB. In dual-unitary circuits the bath can be abstracted by identifying a transfer matrix along the spatial direction. For chaotic dual-unitary circuits this transfer matrix is contracting and has a nontrivial leading eigenvector with unit eigenvalue, such that the repeated application of the transfer matrix in the limit of infinite system size can be replaced by a projection on the leading eigenvector \cite{piroli_exact_2020}. Crucially, this leading (left and right) eigenvector can be directly constructed using unitarity along the spatial direction. Using the identities from Fig.~\ref{fig:circuits}, we find that
\begin{align}
\vcenter{\hbox{\includegraphics[width=0.48\linewidth]{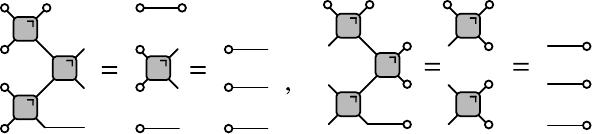}}}\,.
\end{align}
When introducing projective measurements in a UEB, this state remains an eigenstate because our choice of measurement basis preserves unitarity along the spatial direction. However, due to the disentangling effect of measurements in the temporal direction, measurements typically introduce additional leading eigenvectors. The transfer matrix can still be replaced by its projector, since these eigenstates depend on both the position of the measurement and the measurement outcome, such that the above eigenvector is the only common eigenvector. This argument no longer holds in the limit where $p=1$ and all gates consist of projective measurements. In this case the transfer matrices are unitary, such that all eigenvectors have a leading eigenvalue. Still, in this limit the reduced density matrix is a projector purely determined by the measurements at the final time step, and the subsystem is not entangled with the remainder of the system, such that the expression from the main text also holds.

Replacing the bath by this projector and further simplifying the diagram using the spatial unitarity returns
\begin{align}
\vcenter{\hbox{\includegraphics[width=0.55\linewidth]{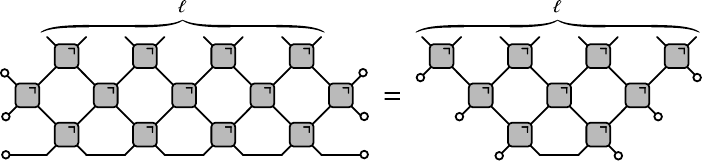}}}\,.
\end{align}
Now considering the case where every gate is either a dual-unitary gate or a projector with probabilities $(1-p)$ and $p$ respectively, it is useful to illustrate specific realizations for e.g. $\ell=8$. Here we consider the limit $t\geq \ell/2$ where the system is thermalized, but the argument immediately extends to shorter times. 
\begin{align}\label{eq:sm:examples}
\vcenter{\hbox{\includegraphics[width=0.85\linewidth]{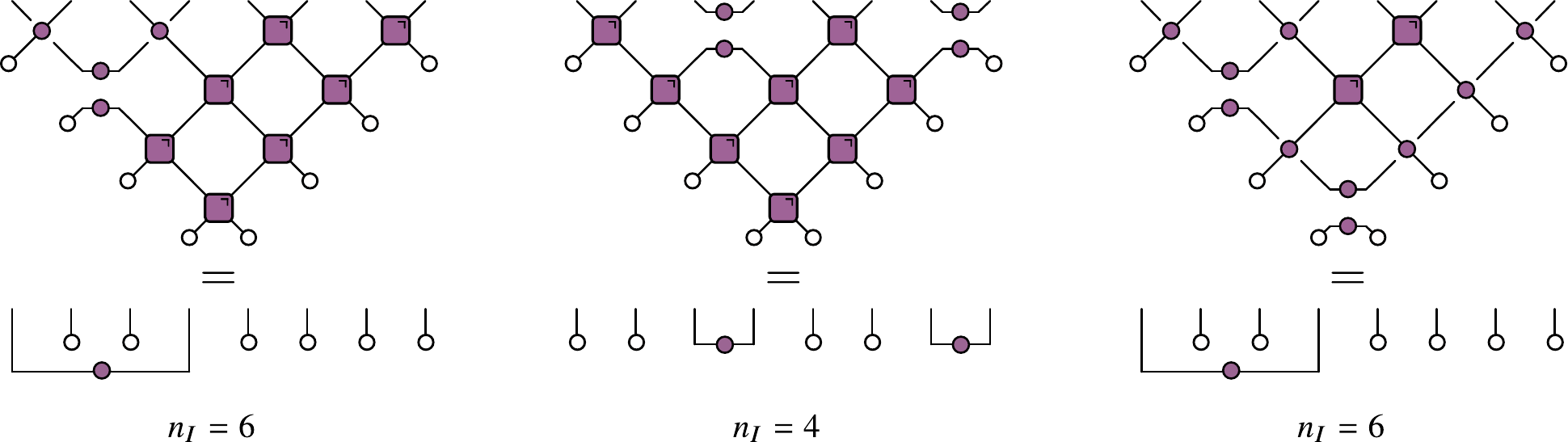}}}
\end{align}
It is clear that every circuit constructed in this way returns a reduced density matrix that acts as the identity $\mathbbm{1}/q$ on $n_{I}$ sites and acts as a projector on the remaining $n_P = \ell-n_{I}$ sites. The spectrum of every density matrix can then immediately be determined as consisting of $q^{n_I}$ eigenvalues equal to $1/q^{n_I}$ and the remaining eigenvalues equal to $0$. The entanglement entropies (both R\'enyi and von Neumann) then follow as $S_A = n_I \ln(q)$. Note that this property again depends on the restricted gates on the light cones of measurements, since otherwise it is not guaranteed that the RDM simplifies in the above way.

The entropy depends on the choice of gates vs. measurements but not on the specific gates and measurement outcomes, such that we can replace every dual-unitary gate by a swap gate and every measurement by a projection on the identity without changing the entropy. The total number of nonzero eigenvalues can now be counted by introducing appropriate boundary conditions,
\begin{align}
\vcenter{\hbox{\includegraphics[width=0.25\linewidth]{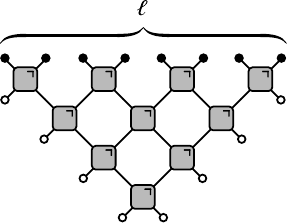}}}\qquad \textrm{with}\qquad\vcenter{\hbox{\includegraphics[width=0.04\linewidth]{ov_bb}}}=\vcenter{\hbox{\includegraphics[width=0.04\linewidth]{ov_ww}}}=1\,,\,\vcenter{\hbox{\includegraphics[width=0.04\linewidth]{ov_bw}}}=1-\epsilon \ln(q)\,.
\end{align}
and where the gray gates are either swap gates or projections on the identity. These diagrams evaluate to $[1-\epsilon \ln(q)]^{n_I} = 1-\epsilon n_I \ln(q)+\mathcal{O}(\epsilon^2)$, as can be illustrated for the examples from Eq.~\eqref{eq:sm:examples} with all measurements replaced by projections on the identity,
\begin{align}
\vcenter{\hbox{\includegraphics[width=0.85\linewidth]{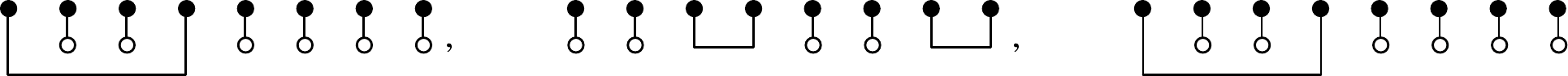}}}\,.
\end{align}
The averaged entropy for a given measurement rate $p$ is obtained by considering a circuit as given in the main text, where every (folded) gate is chosen to be either a projection on the identity or a swap gate with probabilities $p$ and $1-p$ respectively,
\begin{align}
\vcenter{\hbox{\includegraphics[width=0.3\linewidth]{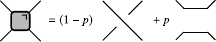}}}\,.
\end{align}

A specific realization of the boundaries can be considered as e.g. $\vcenter{\hbox{\includegraphics[width=0.03\linewidth]{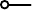}}}=(1,0)$ and $\vcenter{\hbox{\includegraphics[width=0.03\linewidth]{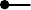}}}=(1-\epsilon \ln(q),\sqrt{2\epsilon\ln(q)-\epsilon^2\ln(q)^2})$. Note that all gates now act on a smaller Hilbert space, but will evaluate to the same result since the above expression only depends on the overlaps. Expanding the full expression in $\sqrt{\epsilon}$, the first-order contribution vanishes and the second order $\mathcal{O}(\epsilon)$ returns
\begin{align}\label{eq:sm:der}
\vcenter{\hbox{\includegraphics[width=0.6\linewidth]{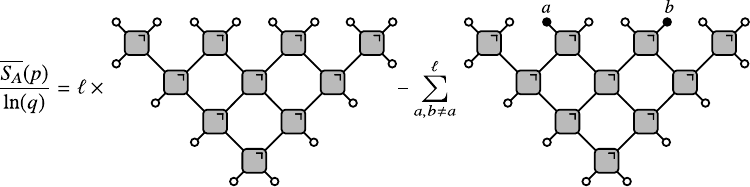}}},
\end{align}
where we have redefined the two states to be orthonormal, $\vcenter{\hbox{\includegraphics[width=0.03\linewidth]{state_w}}}=(1,0)$ and $\vcenter{\hbox{\includegraphics[width=0.03\linewidth]{state_b}}}=(0,1)$. The first diagram can be directly evaluated as $1$, whereas the second diagram can be interpreted as a six-vertex model from statistical mechanics. Since  $\vcenter{\hbox{\includegraphics[width=0.04\linewidth]{ov_bw}}}=0$ the first diagram can be expanded in terms of paths connecting $a$ to $b$, e.g. for $b-a=7$,
\begin{align}
\vcenter{\hbox{\includegraphics[width=\linewidth]{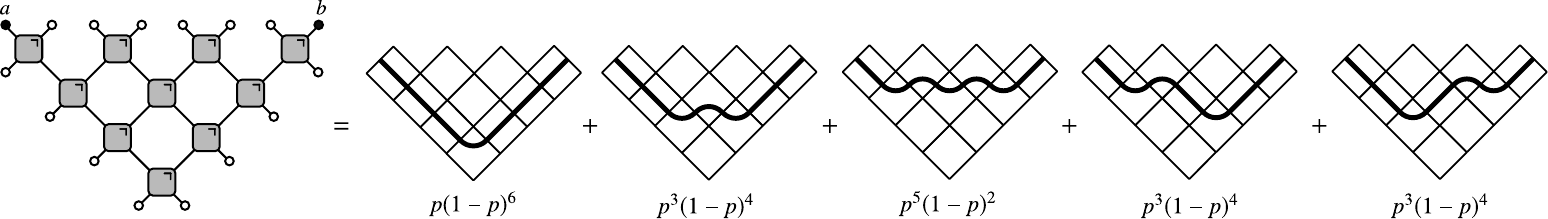}}}\,,
\end{align}
with every path appropriately weighted by the probability of the required swap gates and projections, i.e.
\begin{align}
\vcenter{\hbox{\includegraphics[width=0.35\linewidth]{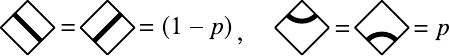}}}\,.
\end{align}
These paths correspond to Dyck paths of order $n=(b-a-1)/2$, and the total number of such paths is given by the Catalan number $C_n = \frac{1}{n+1}{2n \choose n}$ \cite{stanley_catalan_2015}. The weight of each path is fixed by the number of peaks, and the total number of such paths with $k$ peaks is given by the Narayana numbers $N(n,k) = \frac{1}{n} {n \choose k} {n\choose k-1}$. The total number of paths can be recovered from the identity $\sum_{k=1}^n N(n,k) = C_n$.
Additionally, for $b-a=1$ and both located on the same gate only a single path is possible and the diagram returns a weight $p$. Taking these together, we find that the above diagram \eqref{eq:sm:der} evaluates to
\begin{align}\label{eq:sm:comb_SA}
\overline{S_A}(p)/\ln(q) = (1-p)\ell - 2\sum_{n=1}^{\ell/2-1} \left(\frac{\ell}{2}-n\right)\sum_{k=1}^n N(n,k) p^{2k-1}(1-p)^{2(n-k+1)}\,.
\end{align}
This equation alternatively allows a representation in terms of the Gaussian hypergeometric function by identifying the correct combinatorial factors, leading to
\begin{align}
p\overline{S_A}(p)/\ln(q)  = p(1-p)\ell-2 p^2\sum_{n=1}^{\ell/2-1} \left(\frac{\ell}{2}-n\right)(1-p)^{2n} \,_2F_1\left(-n+1,-n,2;z={p^2}/{(1-p)^2}\right) \,,
\end{align}
which can in turn be written in terms of associated Legendre polynomials, with $z=p^2/(1-p)^2$,
\begin{align}
p\overline{S_A}(p)/\ln(q)  = p(1-p)\ell-2p(1-p)\sum_{n=1}^{\ell/2-1}\left(\frac{\ell}{2}-n\right)(1-2p)^{n} P_n^{-1}\left(\frac{1+z}{1-z}\right)\,.
\end{align}

\section{Analytic derivation of the entanglement entropy}
The summation of Eq.~\eqref{eq:sm:comb_SA} can be explicitly evaluated to return a closed-form expression from which the scaling with subsystem size can be established. We will first illustrate the derivation for the symmetric point $p=1/2$ before considering the general problem.
For measurement rate $p=1/2$ all paths have equal weight and the expression for the averaged entanglement entropy reduces to a purely combinatorial one
\begin{align}
\overline{S_A}(p=1/2) = \ln(q)\left[\frac{\ell}{2} -\sum_{n=1}^{\ell/2-1} \left(\frac{\ell}{2}-n\right)\sum_{k=1}^n N(n,k)\, 4^{-n} \right] = \ln(q)\left[\frac{\ell}{2} -\sum_{n=1}^{\ell/2-1} \left(\frac{\ell}{2}-n\right) \, 4^{-n}C_n \right],
\end{align}
with $C_n$ the Catalan numbers, $C_n = \frac{1}{n+1}{2n \choose n}$, counting the total number of Dyck paths of order $n$ \cite{stanley_catalan_2015}. The remaining summation can be explicitly evaluated by using two identities for the Catalan numbers,
\begin{align}
\sum_{n=1}^{\ell/2-1} 4^{-n}C_n = 1- \frac{1}{2^{\ell-1}}{\ell \choose \ell/2}\,,\qquad \sum_{n=1}^{\ell/2-1}  n\, 4^{-n} C_n = -2+ \frac{\ell/2+1}{2^{\ell-1}}{\ell \choose \ell/2}\,.
\end{align}
Both can be derived from the integral representation of the Catalan numbers~\cite{stanley_catalan_2015},
\begin{align}
C_n = \frac{1}{2\pi} \int_{0}^4 dx\, x^n \sqrt{\frac{4-x}{x}} \,,
\end{align}
taking the summation inside the integral and evaluating the geometric power series for $(x/4)$. With $m=\ell/2$ this procedure returns
\begin{align}
&\sum_{n=1}^{m-1} 4^{-n}C_n  = \frac{2}{\pi} \int_{0}^1 dt\,\frac{t-t^m}{1-t}\sqrt{\frac{1-t}{t}} = \frac{2}{\pi}\left[\frac{\pi}{2}-\frac{\sqrt{\pi}\,\Gamma(m+1/2)}{\Gamma(m+1)}\right] = 1- \frac{1}{2^{2m-1}}{2m \choose m}\,,\\
&\sum_{n=1}^{m-1} n\, 4^{-n}C_n  = \frac{2}{\pi} \int_{0}^1 dt\,\frac{(m-1)t^{m+1}-m t^{m}+t}{(1-t)^2}\sqrt{\frac{1-t}{t}} = \frac{2}{\pi}\left[-\pi+(m+1)\frac{\sqrt{\pi}\,\Gamma(m+1/2)}{\Gamma(m+1)}\right] = -2+ \frac{m+1}{2^{2m-1}}{2m \choose m}\,, 
\end{align}
where we have made the substitution $t=x/4$. All resulting integrals define beta functions, which can in turn be expressed as gamma functions and explicitly evaluated. The first identity appears in the Addenda for Ref.~\cite{stanley_catalan_2015}, but we haven't encountered the second identity in the literature \footnote{We acknowledge Richard Stanley for providing an alternative proof starting from the generating function of the Catalan numbers.}.
Using these identities, we recover the result from the main text
\begin{align}\label{eq:sm:SA_sym}
\overline{S_A}(p=1/2) = 2 \ln(q)\left[\frac{\ell+1}{2^{\ell-1}}{\ell \choose \ell/2}-1\right]\,.
\end{align}

Away from the symmetric point the above derivation can be extended by considering associated Legendre polynomials and their integral representation,
\begin{align}
P_n^{-1}\left(x\right) = \frac{\sqrt{x^2-1}}{\pi}\int_{-1}^1 dt\,\sqrt{1-t^2}\left(x+t\sqrt{x^2-1}\right)^{n-1}\,.
\end{align}
Plugging in $x=(1+z)/(1-z)$ with $z=p^2/(1-p)^2$, we can write
\begin{align}
\sum_{n=1}^{m-1} (m-n)\,(1-2p)^n P_n^{-1}\left(\frac{1-2p(1-p)}{1-2p}\right) = \sum_{n=1}^{m-1}(m-n)\frac{2p(1-p)}{\pi}\int_{-1}^1 dt\,\sqrt{1-t^2}\left(1-2p(1-p)(1-t)\right)^{n-1}\,,
\end{align}
where the geometric summation can again be explicitly performed to return, with $\tilde{z}=4p(1-p)$,
\begin{align}
\sum_{n=1}^{m-1} (m-n)\,(1-2p)^n P_n^{-1}\left(\frac{1-2p(1-p)}{1-2p}\right) = \frac{2}{\pi \tilde{z}}\int_{-1}^1 dt \,\frac{\sqrt{1-t^2}}{(1-t)^2} 
\left[(1-\tilde{z}(1-t)/2)^m+m \tilde{z}(1-t)/2-1\right]\,.
\end{align}
Performing a change of variables to $s=(1-t)/2$, the above expression reduces to
\begin{align}
\frac{2}{\pi \tilde{z}}\int_{0}^1 ds\, s^{-3/2}(1-s)^{1/2}\left[(1-\tilde{z}s)^m+m\tilde{z}s-1\right]\,.
\end{align}
This expression can be identified as an integral representation for the Gaussian hypergeometric function $\,_2F_1$, returning
\begin{align}
m+\frac{2}{\tilde{z}}\left[1-\,_2F_1(-m,-1/2,1;\tilde{z})\right]\,.
\end{align}
The full expression for the average entanglement entropy then follows as
\begin{align}
\frac{p\overline{S_A}(p)}{\ln(q)} = p(1-p)\ell-2p(1-p)\sum_{n=1}^{\ell/2-1}(1-2p)^{n} P_n^{-1}\left(\frac{1+z}{1-z}\right)= \,_2F_1(-\ell/2,-1/2,1;4p(1-p))-1\,.
\end{align}
At $p=1/2$ the hypergeometric function is evaluated at the special point $z=1$, where Gauss' summation theorem returns the previously obtained expression \eqref{eq:sm:SA_sym}. The hypergeometric function has known asymptotics when one or more of the parameters is large (see e.g. \cite{bateman_manuscript_project_higher_1953}), and for large $\ell$ we have
\begin{align}
\,_2F_1(-\ell/2,-1/2,1;z=4p(1-p)) \approx \sqrt{\frac{8p(1-p)\,\ell}{\pi}}\,.
\end{align}
For $p=1/2$ this approximation returns Stirling's approximation. 


\section{Entanglement growth}
At short times $t \ll \ell/2$ the expression for the averaged entanglement can be written as
\begin{align}
\vcenter{\hbox{\includegraphics[width=0.5\linewidth]{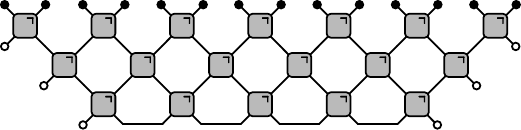}}}\,,
\end{align}
with again $\vcenter{\hbox{\includegraphics[width=0.04\linewidth]{ov_bb}}}=\vcenter{\hbox{\includegraphics[width=0.04\linewidth]{ov_ww}}}=1$ and $\vcenter{\hbox{\includegraphics[width=0.04\linewidth]{ov_bw}}}=1-\epsilon \ln(q)$, and it is possible to identify a transfer matrix along the spatial direction as
\begin{align}
\vcenter{\hbox{\includegraphics[width=0.15\linewidth]{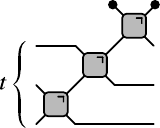}}}\,\qquad \textrm{with} \qquad\vcenter{\hbox{\includegraphics[width=0.28\linewidth]{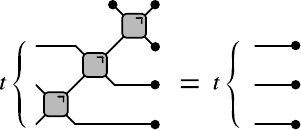}}}\,,
\end{align}
where a (left and right) eigenstate with eigenvalue $1$ can be identified by noting that the gate acts either as the identity or as a swap gate along the spatial direction. It can be easily checked that there are no other leading eigenvalues, such that the repeated application of this transfer matrix can be replaced by a projection on this eigenstate, and the above expression evaluates to
\begin{align}
\left(\vcenter{\hbox{\includegraphics[width=0.04\linewidth]{ov_bw}}}\right)^{2t}=\left(1-\epsilon \ln(q)\right)^{2t} \approx 1- \epsilon\, 2t\,\ln(q)\,.
\end{align}
Corrections on this entanglement growth arise from subleading eigenvalues and are hence exponentially suppressed as $\ell$ increases (for fixed $t< \ell/2$).

For $t=1$ the total expression is particularly simple, 
\begin{align}
\vcenter{\hbox{\includegraphics[width=0.35\linewidth]{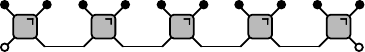}}}\,,
\end{align}
where the only way to get an overlap $\vcenter{\hbox{\includegraphics[width=0.04\linewidth]{ov_ww}}}$ is to choose all gates to be projections, and otherwise the diagram will evaluate to $(\vcenter{\hbox{\includegraphics[width=0.04\linewidth]{ov_bw}}})^2$. The diagram evaluates to
\begin{align}
p^{\ell/2}(\vcenter{\hbox{\includegraphics[width=0.04\linewidth]{ov_ww}}})+(1-p^{\ell/2})(\vcenter{\hbox{\includegraphics[width=0.04\linewidth]{ov_bw}}})^2 = p^{\ell/2}+(1-p^{\ell/2})(1-\epsilon \ln(q))^2 = 1 - 2 \epsilon (1-p^{\ell/2}) \ln(q)+\mathcal{O}(\epsilon^2)\,.
\end{align}

\end{widetext}
\end{document}